\pdfoutput=1

\documentclass[11pt,dvipsnames]{article}

\usepackage[final]{acl}

\usepackage{times}
\usepackage{latexsym}

\usepackage[T1]{fontenc}

\usepackage[utf8]{inputenc}

\usepackage{microtype}

\usepackage{inconsolata}

\usepackage{graphicx}

\usepackage{algorithm}
\usepackage{algorithmic}
\usepackage{amsmath}
\usepackage{amssymb}
\usepackage{booktabs}
\usepackage{xcolor}
\usepackage{tabularx} 
\usepackage{multirow}
\usepackage{array} 
\usepackage{makecell}

\title{DiVISe: Direct Visual-Input Speech Synthesis \\
Preserving Speaker Characteristics And Intelligibility}

\author{Yifan Liu \\
  Shanghai Jiao Tong University \\
  \texttt{yifan.liu@sjtu.edu.cn} \And
  Yu Fang \\
  ShanghaiTech University \\
  \texttt{fangyu2022@shanghaitech.edu.cn} \\\AND
  Zhouhan Lin \\
  Shanghai Jiao Tong University \\
  \texttt{lin.zhouhan@gmail.com}
  }

\newcommand{\lrsbasesetting}{30h }
\newcommand{\lrsfullsetting}{433h }
\newcommand{\lrstwofullsetting}{224h }

\newcommand{\unithifigan}{Unit-HiFiGAN}
\begin{document}
\maketitle
\begin{abstract}
Video-to-speech (V2S) synthesis, the task of generating speech directly from silent video input, is inherently more challenging than other speech synthesis tasks due to the need to accurately reconstruct both speech content and speaker characteristics from visual cues alone.
Recently, audio-visual pre-training has eliminated the need for additional acoustic hints in V2S, which previous methods often relied on to ensure training convergence. However, even with pre-training, existing methods continue to face challenges in achieving a balance between acoustic intelligibility and the preservation of speaker-specific characteristics.
We analyzed this limitation and were motivated to introduce DiVISe (\textbf{Di}rect \textbf{V}isual-\textbf{I}nput \textbf{S}peech Synth\textbf{e}sis), an end-to-end V2S model that predicts Mel-spectrograms directly from video frames alone.
Despite not taking any acoustic hints,
DiVISe effectively preserves speaker characteristics in the generated audio, and achieves superior performance on both objective and subjective metrics across the LRS2 and LRS3 datasets. Our results demonstrate that DiVISe not only outperforms existing V2S models in acoustic intelligibility but also scales more effectively with increased data and model parameters\footnote{Code and weights can be found at \url{https://github.com/PussyCat0700/DiVISe}.}.
\end{abstract}

\section{Introduction}
Video-to-speech (V2S) synthesis aims to generate speech from silent speaker videos. Despite its challenging nature, V2S remains valuable because it conveys richer speaker information beyond language, such as timbre, pitch, tone, and mood, which can be easier to comprehend compared to text transcribed by visual speech recognition.
V2S holds immense potential in applications such as restoring speech for individuals with speech impairments or enabling communication in noisy environments \citep{prajwalLip2wav}.

Since V2S needs to generate waveforms that accurately reconstruct both speech content and timbre from silent lip movements, it is considered challenging. The lack of prior knowledge about the speech corresponding to the speaker's video clips can confuse the model during training and lead to poor synthesis results \citep{revisewnhu, choi2023diffv2s}.
Most existing V2S approaches \cite{mira2022svts, Hegde2023TowardsAccLR, Kim2023MultiTaskL2S} require an additional speaker embedding derived from extra audio (e.g., random audio clips from the same speaker) to provide speaker characteristics as prior knowledge.
However, these models often face challenges in practical scenarios, as speaker identity or voice samples are typically unavailable, especially during inference.

Recent advances in V2S aim to eliminate reliance on speaker embeddings. A major breakthrough is offered by ReVISE \cite{revisewnhu}, which leverages AV-HuBERT \citep{Shi2022AVHuBERT} as its visual backbone, pre-trained on a combination of audio-visual modalities. By utilizing the prior knowledge encoded in this visual backbone, ReVISE generates high-quality audio without requiring speaker embeddings. To convert the visual backbone's output into waveforms, ReVISE uses a separate vocoder, Unit-HiFiGAN, based on \citet{lee-etal-2022-direct}, which takes discrete acoustic units clustered from an audio encoder \citep{hubert} as input. However, the generated audio lacks speaker-specific characteristics. Similarly, DiffV2S \cite{choi2023diffv2s} employs pre-training with AV-HuBERT \citep{Shi2022AVHuBERT} and introduces a speaker embedding-conditioned diffusion model to generate Mel-spectrograms from silent videos. Although speaker embeddings are not required during inference, they are still necessary during a separate training phase before training the diffusion model.

In this work, we propose \textbf{Di}rect \textbf{V}sual-\textbf{I}nput \textbf{S}peech Synth\textbf{e}sis (DiVISe), an end-to-end, speaker-embedding-free approach that leverages audio-visual pre-training \citep{Shi2022AVHuBERT} to directly predict Mel-spectrograms from video frames. Thanks to its end-to-end design and simple training process, DiVISe is easy to scale. Our motivation comes from analyzing the Unit-HiFiGAN vocoder used by ReVISE \cite{revisewnhu}, which reveals that reliance on acoustic units adversely affects the speaker similarity of the generated audio. Notably, DiVISe achieves superior objective and subjective evaluation results on the LRS2 and LRS3 datasets. The main contributions of this work are as follows:
\begin{itemize}
    \item Our analysis of ReVISE \citep{revisewnhu} revealed that the loss of speaker characteristics is primarily due to its choice of vocoder. The unit-based vocoder struggles to preserve speaker characteristics, whereas vocoders based on Mel-spectrograms perform significantly better in retaining them.
    \item We introduced DiVISe, an end-to-end V2S approach that predicts Mel-spectrograms. By leveraging audio-visual pre-training and an effective vocoder, DiVISe is not only efficient to train but also excels at preserving speaker characteristics.
    \item DiVISe excels in generating high-quality audio. Even with visual-only input, it outperforms existing methods that rely on additional audio inputs in both objective and subjective evaluations. Moreover, DiVISe's performance scales efficiently with increased data or model parameters.  
    

    
\end{itemize}

\section{Related Work}

\begin{figure*}[t]
    \centering
    \includegraphics[width=\textwidth]{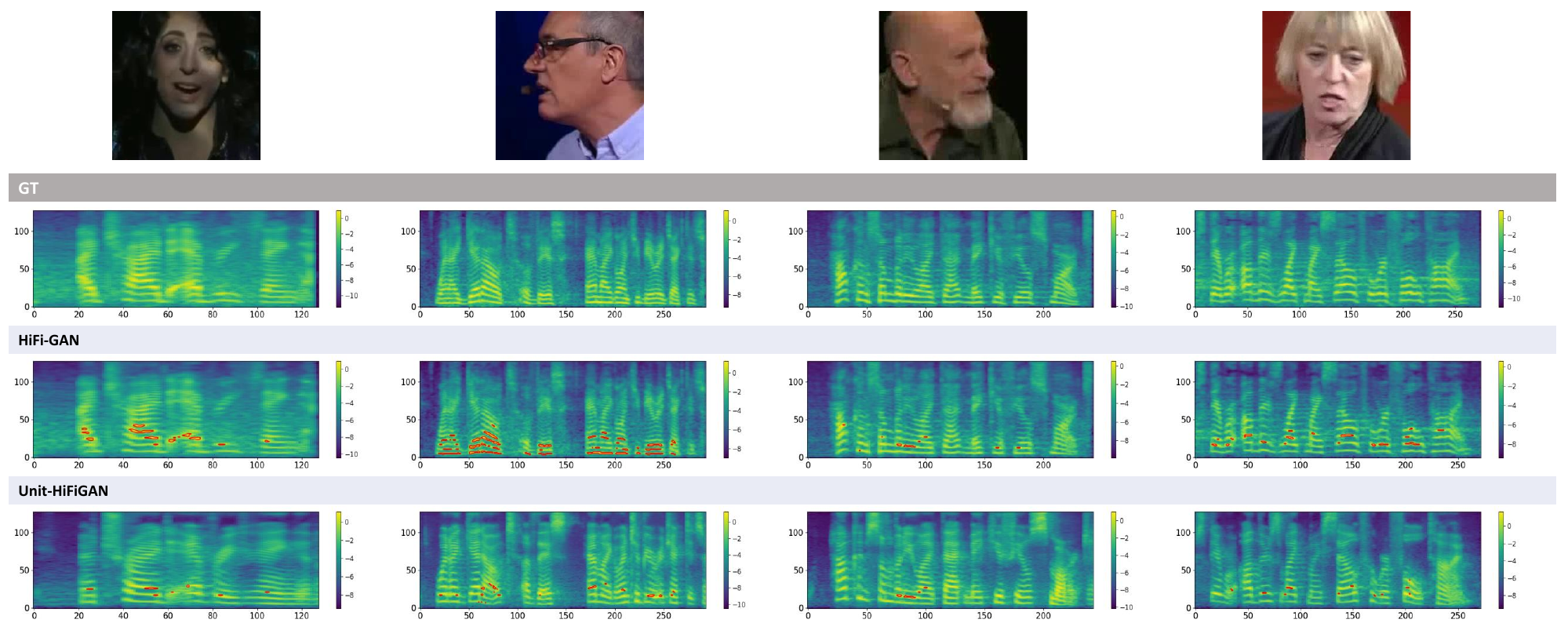}
    \caption{Comparison of the Mel-spectrograms of audios generated by HiFi-GAN (middle) and Unit-HiFiGAN (bottom). To highlight the active speaking regions that overlap the most with the ground truth speech (top), regions with smaller numerical differences from the ground truth are marked in red.}
    \label{fig:mels_of_ids}
\end{figure*}

\subsection{Speech Synthesis With Vocoders}

Generating audio waveforms directly in speech synthesis is challenging because of the complexity of audio signals; therefore, vocoders are employed to convert acoustic features, such as Mel-spectrograms or hidden units, into audio waveforms. Mel-spectrograms are often used as an intermediate audio feature for prediction targets due to their interpretability and computational efficiency in the continuous frequency domain.
\citet{parallelWavegan_PWG} proposed Parallel WaveGAN (PWG), which utilizes multi-resolution loss in GAN training. BigVGAN \cite{lee2023bigvgan} is trained on a large-scale dataset and successfully improves audio quality. \citet{kong2020hifigan} introduced HiFi-GAN, which employs a combination of multi-period and multi-scale discriminators to differentiate between the generated audio waveform from the Mel-spectrogram and the original waveform during training. Alternatively, non-parametric methods like the Griffin-Lim algorithm \cite{griffinlim} provide a less accurate estimation of the waveform from the Mel-spectrogram.

\citet{Polyak2021SpeechRF} proposed a new speech re-synthesis model that successfully converts pre-trained HuBERT units back into audible speech. This approach involves adapting HiFi-GAN \citep{kong2020hifigan} to enable direct waveform generation from SSL (Self-Supervised Learning) units. Later, \citet{lee-etal-2022-direct} discovered that when trained on LJSpeech \cite{ljspeech17}, reliance on speaker embeddings and fundamental frequency can be reduced, thereby streamlining the model's input requirements. Following this insight, \citet{revisewnhu} proposed the ReVISE model, which combines pre-trained AV-HuBERT \cite{Shi2022AVHuBERT} with an upsampled linear projection module that generates SSL unit predictions. Similar to \citet{lee-etal-2022-direct}, the SSL units are labeled by a pre-trained HuBERT model and converted back to the original waveform with the help of the Unit HiFi-GAN model \cite{revisewnhu}.

\subsection{Video-To-Speech Synthesis}

Early works of V2S requires speaker embedding as input to ensure acoustic intelligibility. \citet{prajwalLip2wav} developed Lip2Wav
and explored training the model on a mixture of speakers with the help of speaker embedding similar to \citet{jia2019transfer}. However, the multi-speaker capability of Lip2Wav heavily depended on these speaker embeddings, without which the model would suffer performance degradation.
Inspired by Lip2Wav \cite{prajwalLip2wav}, SVTS \cite{mira2022svts} proposed a V2S system built with ResNet18 \cite{resnet18} and Conformer \cite{Gulati2020ConformerCT}, and trained their method on a combination of 1759 hours of audio-visual data comprising LRS3 \cite{Afouras2018LRS3TEDAL} and VoxCeleb2 \cite{VoxCeleb2Chung_2018}. However, to adapt their approach to a multi-speaker context, SVTS also relied on speaker embeddings from extra audio clips during both training and inference phases. Following SVTS, \citet{Kim2023MultiTaskL2S, Hegde2023TowardsAccLR} incorporated additional textual information and speaker embeddings to further enhance V2S performance. \citet{Kim2023MultiTaskL2S} based their V2S model on SVTS and proposed to align the intermediate feature of their V2S model with the textual transcription of the speech. \citet{Hegde2023TowardsAccLR} integrated a lip-to-text model in their training pipeline along with a modified text-to-speech (TTS) model that utilizes speaker embeddings to produce well-synchronized audio waveforms. Despite the progress these approaches represent in V2S synthesis, they have not eliminated the reliance on additional inputs, limiting the practical uses of these V2S methods.


\section{Preliminaries: Speaker Characteristics Preservation In Vocoders}

In Video-to-Speech synthesis, effective speech recovery requires the simultaneous preservation of both content information and speaker characteristics. \citet{lee-etal-2022-direct,revisewnhu} proposed using acoustic units as vocoder input for speech recovery, successfully encoding phonetic information to reconstruct speech content. However, \citet{revisewnhu} found that these units fail to preserve speaker characteristics and cannot generate sound that matches the speaker's identity.

In this work, we first illustrate the limitations of the unit-based vocoder in speech synthesis through a straightforward visual comparison of Mel-spectrograms, highlighting the differences in generated speech within the frequency domain. In Figure \ref{fig:mels_of_ids}, we compare the Mel-spectrograms of speech generated by the unit-based vocoder, Unit-HiFiGAN \citep{revisewnhu}, and the mel-based vocoder, HiFi-GAN \citep{kong2020hifigan}\footnote{For visualization, frequency components with values greater than 0.0 and differences of less than 0.8 from the ground truth are highlighted in red.}. The unit-based vocoder shows little resemblance to the original speech in the frequency domain, while the mel-based vocoder aligns closely with the original frequency components, making it more effective at preserving speaker characteristics. This indicates that the speech generated by the mel-based vocoder matches the original speech more closely than that produced by the unit-based vocoder.

From a quantitative perspective, Table \ref{tab:speaker_verification1} compares the performance of the two vocoders. The unit-based vocoder underperforms across all metrics, particularly in preserving speaker characteristics (SECS and EER scores), indicating that it is not well-suited for speech synthesis. Furthermore, intelligibility metrics (PESQ, STOI, ESTOI) and content accuracy (WER) reinforce the conclusion that the mel-based vocoder is a superior choice for speech synthesis. These results confirm that the mel-based vocoder recovers speaker characteristics more accurately than the unit-based vocoder.

Since the V2S task naturally allows for inferring speaker characteristics from visual cues, and given that ReVISE \citep{revisewnhu} loses speaker identity due to its unit-based vocoder, we are motivated to use Mel-spectrograms as prediction targets in V2S synthesis instead of acoustic units. By employing the mel-based vocoder for speech recovery, we can better capture visual speaker characteristics and reflect speaker variations in the synthesized speech.

\begin{table}[t]
\centering
\small
\setlength{\tabcolsep}{0.5mm}{%
\begin{tabular}{lccccc}
\toprule
\multirow{2}{*}{\textbf{Method}} & \multicolumn{2}{c}{Match} & \multicolumn{3}{c}{Intell.} \\
&\textbf{SECS}$\uparrow$&\textbf{EER}$\downarrow$&\textbf{PESQ}$\uparrow$&\textbf{ESTOI}$\uparrow$&\textbf{WER}$\downarrow$\\
\hline
Unit-HiFiGAN & 0.555 & 40.52 
& 1.17 & 0.45 & 14.28 \\
HiFi-GAN & \textbf{0.894} & \textbf{22.96} 
& \textbf{2.91} & \textbf{0.84} & \textbf{7.63} \\
\bottomrule
\end{tabular}%
}
\caption{Audio synthesis performance of HiFi-GAN and Unit-HiFiGAN on speaker matching and intelligibility.}
\label{tab:speaker_verification1}
\end{table}

\section{Methodology}

\begin{figure*}[t]
    \centering
    \includegraphics[width=\textwidth]{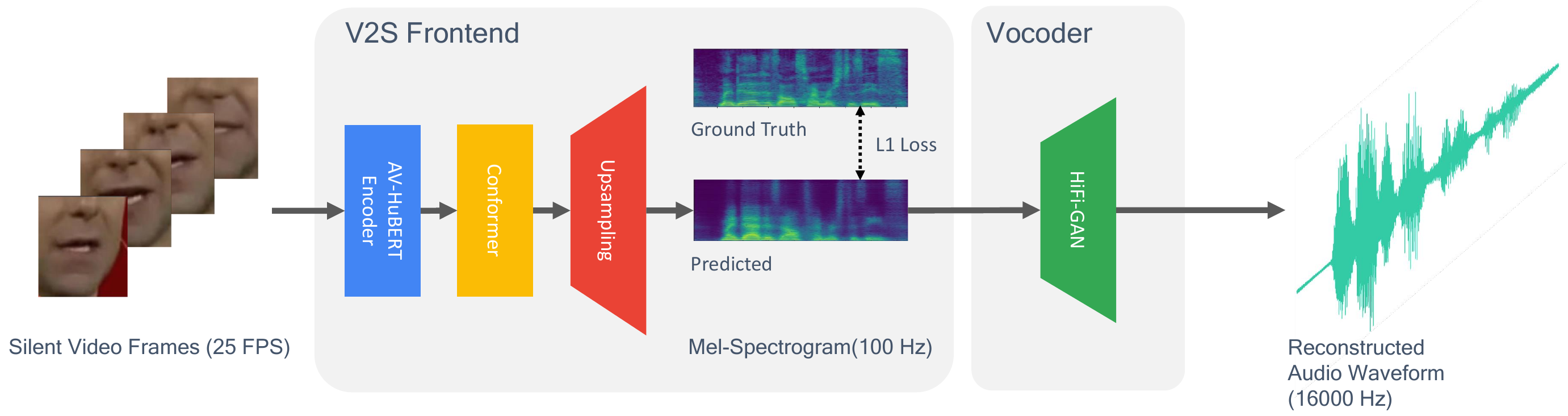}
    \caption{Overview of the DiVISe framework. The V2S frontend maps visual features to Mel-spectrograms, which are then converted into waveforms by the vocoder. During training, DiVISe uses only visual input to generate Mel-spectrograms. During evaluation, the generated Mel-spectrograms are transformed back into audio waveforms.}
    \label{fig:model_architecture}
\end{figure*}

The DiVISe framework, shown in Figure \ref{fig:model_architecture}, comprises two main steps: first, the V2S frontend generates Mel-spectrograms from silent video frames; then, a neural vocoder converts these Mel-spectrograms into audio waveforms. Because the vocoder is trained solely on audio data, the V2S frontend serves as the key component in the overall training pipeline.

\subsection{Model Architecture}\label{sec:model_architecture}

The V2S frontend comprises a pre-trained AV-HuBERT encoder \citep{Shi2022AVHuBERT} integrated with a conformer module. In line with the V2S methodology of \citet{choi2023diffv2s}, we employ HiFi-GAN \citep{kong2020hifigan} as our vocoder. To ensure synchronization between the output length of the V2S frontend and the vocoder, we utilize linear reshaping \citep{Shi2022AVHuBERT}, which upsamples the temporal resolution to align with the Mel-spectrogram. A detailed analysis of the model layout can be found in Section \ref{sec:model_layout}.

The AV-HuBERT encoder comprises a modified ResNet-18 \cite{resnet18} and a transformer encoder \citep{transformer_attentionisalluneed}. In our configuration, the AV-HuBERT model features a 12-layer transformer with 1024 embedding dimensions and 16 attention heads per layer. Following the AV-HuBERT encoder, we integrate a conformer module \citep{Gulati2020ConformerCT} to capture temporal dependencies. This integration aligns the encoder's outputs with the target Mel-spectrogram, enabling a smoother transition between stages. The conformer module utilized in our study consists of four conformer blocks, each equipped with a 256-dimensional attention mechanism and four attention heads.

\subsection{Method Description}

We define the problem of V2S synthesis as follows. The model is given a sequence of video frames $\mathbf{v} = [\mathbf{v}_1, \mathbf{v}_2, \ldots, \mathbf{v}_m] \in \mathbb{R}^{T_\mathbf{v} \times H \times W \times 1}$, where $T_\mathbf{v}$ is the temporal dimension, and $H$, $W$ are the height and width of the frames, respectively. Our objective is to transform the input visual frames into a corresponding sequence of audio waveforms $\mathbf{a} = [\mathbf{a}_1, \mathbf{a}_2, \ldots, \mathbf{a}_n] \in \mathbb{R}^{T_\mathbf{a}}$, where $T_\mathbf{a}$ is the length of the audio sequence. Specific configurations for preprocessing and parameters in this section are elaborated in Section \ref{sec:training_details}.

\subsubsection{V2S frontend}
This section outlines the structure, training, and evaluation of the V2S frontend, which converts visual input into Mel-spectrograms.
\paragraph{Visual Backbone}
We utilize the AV-HuBERT encoder to process video frames $\mathbf{v}$, generating a sequence of embeddings $\mathbf{e_v} \in \mathbb{R}^{T_\mathbf{v} \times C}$, where $C$ represents the feature dimension. This encoder leverages audio-visual pre-training to provide the necessary prior knowledge for our V2S pipeline.
\paragraph{Upsampling}
To align the temporal resolution of the visual embeddings $\mathbf{e_v}$ with the Mel-spectrogram's sampling rate, we apply linear upsampling, increasing the resolution from $T_\mathbf{v}$ to $kT_\mathbf{v}$. This produces upsampled embeddings $\mathbf{e_m} \in \mathbb{R}^{kT_\mathbf{v} \times C/k}$.

\paragraph{Conformer Module}
The upsampled embeddings, $\mathbf{e_m}$, are then passed through a conformer module, which captures long-range dependencies and refines the temporal structure. The output of the conformer is subsequently projected through a linear layer to produce the final Mel-spectrogram $\tilde{\mathcal{S}} \in \mathbb{R}^{kT_\mathbf{v} \times F}$, where $F$ represents the number of frequency bins in the Mel-spectrogram.

\paragraph{Training and Evaluation}
During training, the generated Mel-spectrogram $\tilde{\mathcal{S}}$ is optimized using L1 loss to closely match the ground truth Mel-spectrogram $\mathcal{S} \in \mathbb{R}^{kT_\mathbf{v} \times F}$, ensuring the acoustic accuracy of the generated audio:
\begin{equation}
L_{\text{1}} = \Vert\mathcal{S} - \tilde{\mathcal{S}}\Vert_1. 
\end{equation}

For evaluation, $\tilde{\mathcal{S}}$ is passed to a downstream neural vocoder $\mathcal{V(\cdot; \phi)}$, which converts it back into a speech waveform $\mathbf{w} = \mathcal{V}(\tilde{\mathcal{S}}; \phi), \mathbf{w} \in \mathbb{R}^{T_\mathbf{a}}$. No gradient is propagated back to the V2S frontend during this process. The neural vocoder parameters, denoted as $\phi$, are assumed to be pre-trained and fixed.

\subsubsection{Vocoder}
In this section, we introduce the two-stage training process for the vocoder: pre-training on audio data and fine-tuning on V2S-generated Mel-spectrograms to address distribution mismatch.
\paragraph{Pre-training} During the pre-training phase, the vocoder $\mathcal{V}(\cdot; \phi)$ is trained on an audio-only dataset consisting of ground-truth waveforms $\mathbf{w}$ and their corresponding Mel-spectrograms $\mathcal{S}$, sampled from the distribution $\mathbb{P}_{a}$. The vocoder is optimized to reconstruct $\mathbf{w}$ from $\mathcal{S}$ by maximizing the following expectation:
\begin{equation}
    \max_{\phi} \mathbb{E}_{\mathcal{S} \sim \mathbb{P}_{a}} \left[ \mathcal{V}(\mathcal{S}; \phi) \to \mathbf{w} \right]
\end{equation}

\paragraph{Fine-tuning}

When the V2S frontend generates Mel-spectrograms $\tilde{\mathcal{S}}$, they follow a different data distribution $\mathbb{P}_f$ compared to the vocoder's training distribution $\mathbb{P}_a$, which degrades acoustic perceptual quality \citep{kong2020hifigan}. Therefore, we fine-tune the vocoder on the $\tilde{\mathcal{S}}$ generated by the V2S frontend, adjusting its parameters to handle the distribution shift in waveform reconstruction. 
This fine-tuning process is formalized as:
\begin{equation}
    \max_{\phi} \mathbb{E}_{\tilde{\mathcal{S}} \sim \mathbb{P}_{f}} \left[ \mathcal{V}(\tilde{\mathcal{S}}; \phi) \to \mathbf{w} \right]
\end{equation}
Through this adaptation, the vocoder is optimized to account for the distributional differences between real and generated Mel-spectrograms, ensuring consistent and accurate waveform reconstruction from the V2S frontend outputs.

\section{Experiments}


\subsection{Datasets}
In our experimental setup, we employ two primary datasets for training and evaluating the V2S models: the Lip Reading Sentences 2 (LRS2-BBC) dataset \cite{LRS2Afouras_2022} and the Lip Reading Sentences 3 (LRS3-TED) dataset \cite{Afouras2018LRS3TEDAL}. The LRS3 dataset comprises 433 hours of transcribed English video content, while the LRS2 dataset contains 224 hours. Both datasets feature videos sampled at 25Hz and corresponding audio sampled at 16kHz.
For experiments conducted on both datasets, we define the low-resource condition as \lrsbasesetting. The full-resource settings are denoted as \lrsfullsetting for LRS3 and \lrstwofullsetting for LRS2, respectively.

Additionally, the LJSpeech Dataset \cite{ljspeech17} is utilized for training the neural vocoders. To train HiFi-GAN and Unit-HiFiGAN, we used a resampled version of the LJSpeech Dataset at 16kHz, following \citet{Niekerk2021ACOSoftSpeechUnitsCompare}.
The VoxCeleb2 dataset \cite{VoxCeleb2Chung_2018} is also employed but exclusively for evaluating the EER metric using its test set, as speaker pairs for verification trials can be extracted from this dataset.

\begin{table*}[h!t]
\centering
\small
\textbf{LRS3-TED}
\setlength{\tabcolsep}{1mm}{%
\centering
\begin{tabular*}{\textwidth}{@{\extracolsep{\fill}}lcccccc}
\toprule
\textbf{Method}&\textbf{Data}&\textbf{Vocoder}&\textbf{PESQ}$\uparrow$&\textbf{STOI}$\uparrow$&\textbf{ESTOI}$\uparrow$&\textbf{WER}$\downarrow$\\
\hline
\multicolumn{7}{c}{\textit{Methods taking Visual-only input}} \\
VCA-GAN \cite{Kim2022VCAGAN} & \lrsfullsetting & Griffin-Lim \cite{Kim2023MultiTaskL2S} & \textbf{1.23} & 0.474 & 0.207 &  96.63\\
DiffV2S \cite{choi2023diffv2s} & \lrsfullsetting & HiFi-GAN & -- & -- & 0.284 & 39.2 \\
ReVISE \cite{revisewnhu} & \lrsfullsetting & \unithifigan & -- & -- & -- & \textbf{35.2}  \\
\hline
\multicolumn{7}{c}{\textit{Methods requiring extra information}} \\
VAE-GAN \dag\cite{Hegde_2022VAEGAN} & \lrsfullsetting & - & 0.51 & 0.30 & 0.15 & - \\
AccurateL2S \cite{Hegde2023TowardsAccLR}\ddag & \lrsfullsetting & BigVGAN & 1.39 & 0.58 & 0.37 & - \\
Multi-Task \cite{Kim2023MultiTaskL2S}\ddag & \lrsfullsetting & Griffin-Lim & 1.31 & 0.497 & 0.268 & 66.78 \\
SVTS \cite{mira2022svts}\dag & \lrsfullsetting & PWG & 1.25 & 0.507 & 0.271 & 81.5 \cite{revisewnhu} \\
SVTS \cite{mira2022svts}\dag & 1759h & PWG & 1.26 & 0.530 & 0.313 & 67.4 \cite{revisewnhu} \\
\hline
\multicolumn{7}{c}{\textit{Our Implementation}} \\
ReVISE \citep{revisewnhu} & \lrsbasesetting & \unithifigan & 1.14 & 0.489 & 0.288 & 37.61 \\
ReVISE \citep{revisewnhu} & \lrsfullsetting & \unithifigan & 1.15 & 0.491 & 0.290 & 36.03 \\
DiVISe & \lrsbasesetting & HiFi-GAN & \underline{1.18} & \underline{0.522} & \underline{0.323} & 38.38 \\ 
DiVISe & \lrsfullsetting & HiFi-GAN & \underline{1.17} & \textbf{0.527} & \textbf{0.331} & \underline{35.68} \\
\bottomrule
\end{tabular*}%

}

\textbf{LRS2-BBC}
\setlength{\tabcolsep}{1mm}{%
\begin{tabular*}{\textwidth}{@{\extracolsep{\fill}}lcccccc}
\toprule
\textbf{Method}&\textbf{Data}&\textbf{Vocoder}&\textbf{PESQ}$\uparrow$&\textbf{STOI}$\uparrow$&\textbf{ESTOI}$\uparrow$&\textbf{WER}$\downarrow$\\
\hline
\multicolumn{7}{c}{\textit{Methods taking Visual-only input}} \\
VCA-GAN \cite{Kim2022VCAGAN} & \lrstwofullsetting & Griffin-Lim \cite{Kim2023MultiTaskL2S} & \textbf{1.24} & 0.407 & 0.134 &  109.01\\
DiffV2S \cite{choi2023diffv2s} & \lrstwofullsetting & HiFi-GAN & -- & -- & 0.283 & 52.7 \\
\hline
\multicolumn{7}{c}{\textit{Methods requiring extra information}} \\
VAE-GAN \cite{Hegde_2022VAEGAN}\dag & \lrstwofullsetting & - & 0.60 & 0.34 & 0.17 & - \\
AccurateL2S \cite{Hegde2023TowardsAccLR}\ddag & \lrstwofullsetting & BigVGAN & 1.47 & 0.65 & 0.47 & - \\
Multi-Task \cite{Kim2023MultiTaskL2S}\ddag & \lrstwofullsetting & Griffin-Lim & 1.36 & 0.526 & 0.341 & 60.54 \\
IntelligibleL2S \cite{choi2023intelligible}\dag & \lrstwofullsetting & Hybrid & 1.34 & 0.585 & 0.412 & 35.7 \\
SVTS \cite{mira2022svts}\dag & \lrstwofullsetting & Griffin-Lim \cite{Kim2023MultiTaskL2S} & 1.34 & 0.49 & 0.29 & 93.91 \\
\hline
\multicolumn{7}{c}{\textit{Our Implementation}} \\
ReVISE \citep{revisewnhu} & \lrsbasesetting & \unithifigan & 1.12 & 0.484 & 0.297 & 38.60 \\ 
ReVISE \citep{revisewnhu} & \lrstwofullsetting & \unithifigan & 1.12 & 0.488 & 0.302 & \underline{36.95} \\ 
DiVISe & \lrsbasesetting & HiFi-GAN & \underline{1.23} & \underline{0.521} & \underline{0.342} & 40.74 \\

DiVISe & \lrstwofullsetting & HiFi-GAN & 1.22 & \textbf{0.540} & \textbf{0.367} & \textbf{36.24} \\ 
\bottomrule
\end{tabular*}%
}
\caption{Intelligibility metrics on LRS3 and LRS2 datasets. Top-1/top-2 performances are marked in bold/underlined for methods taking visual input only. Methods marked with \dag require extra speaker embedding in training, while those marked with \ddag require not only audio embedding, but also additional text information. The ASR setting is consistent with \citet{revisewnhu} but may differ in other methods.}
\label{tab:synthesis_results}
\end{table*}

\subsection{Results on Speaker Characteristic Preservation}

\paragraph{Speech Similarity and Speaker Verification}

To test the model's performance in speaker matching, we applied speaker encoder cosine similarity (SECS) \citep{choi2023diffv2s} and equal error rate (EER) as our measurements, both of which are correlated with speaker embeddings extracted from audio. Higher SECS values and lower EER values indicate better performance in speech similarity and speaker verification, respectively. EER is computed only on VoxCeleb2 \citep{VoxCeleb2Chung_2018}, as in \citet{Shi2022LearningAudioVisualSpeakerEmbedding_SCSV}. A detailed explanation of these metrics can be found in Section \ref{sec:metrics}.

Table \ref{tab:diffv2s_secs} presents a comparison of SECS scores on the LRS2 and LRS3 datasets. The results for \citet{mira2022svts, Kim2023MultiTaskL2S, Kim2022VCAGAN, choi2023diffv2s} are taken from the DiffV2S implementation by \citet{choi2023diffv2s}. DiVISe demonstrates leading performance on both datasets, despite not using speaker embeddings during training or evaluation, highlighting its ability to restore speaker characteristics. Additionally, as both DiVISe and DiffV2S \citep{choi2023diffv2s} utilize the pre-trained AV-HuBERT encoder \citep{Shi2022AVHuBERT} as the visual backbone, their SECS scores significantly surpass other methods, indicating the effectiveness of audio-visual pre-training in preserving speaker characteristics.

Table \ref{tab:speaker_verification} shows the comparison on the LRS3 dataset between ReVISE and DiVISe.
Compared to ReVISE, DiVISe performs especially better in speaker characteristics preservation, showing advantage in preserving speaker traits in the synthesized audio. This trend aligns with our observations from the vocoder comparison in Table \ref{tab:speaker_verification1}, signifying mel-based methods consistently outperform unit-based methods especially in speaker matching.

\begin{table}[t]
\centering
\small
\setlength{\tabcolsep}{1mm}{%
\begin{tabular}{cccccc}
\toprule
\multirow{2}{*}{\textbf{Method}} & \multicolumn{2}{c}{Match} & \multicolumn{3}{c}{Intell.} \\
&\textbf{SECS}$\uparrow$&\textbf{EER}$\downarrow$&\textbf{PESQ}$\uparrow$&\textbf{ESTOI}$\uparrow$&\textbf{WER}$\downarrow$\\
\hline
ReVISE & 0.5384 & 41.17 
& 1.15 & 0.290 & 36.03 \\
DiVISe & \textbf{0.6242} & \textbf{28.66}
& \textbf{1.17} & \textbf{0.331} & \textbf{35.68} \\
\bottomrule
\end{tabular}%
}
\caption{Metrics comparison for DiVISe and ReVISE on speaker matching and intelligibility.}
\label{tab:speaker_verification}
\end{table}

\begin{table}[t]
\centering
\small
\setlength{\tabcolsep}{1mm}{%
\begin{tabular}{lcc}
\toprule
\textbf{Method}&\textbf{LRS2}&\textbf{LRS3}\\
\hline
\multicolumn{2}{c}{\textit{w/ audio speaker embedding}} \\
SVTS \citep{mira2022svts} & 0.558 & 0.623 \\
Multi-Task \citep{Kim2023MultiTaskL2S} & 0.525 & 0.549 \\
\hline
\multicolumn{2}{c}{\textit{w/o audio speaker embedding}} \\
VCA-GAN \citep{Kim2022VCAGAN} & 0.453 & 0.445 \\
SVTS \citep{mira2022svts} & 0.499 & 0.543 \\
Multi-Task \citep{Kim2023MultiTaskL2S} & 0.457 & 0.495 \\
DiffV2S \citep{choi2023diffv2s} & \underline{0.581} & \textbf{0.625} \\
\hline
ReVISE \citep{revisewnhu} & 0.521 & 0.538 \\
DiVISe & \textbf{0.609} & \underline{0.624} \\

\bottomrule
\end{tabular}%
}
\caption{SECS comparison of methods. Top-1/top-2 performances are marked in bold/underlined for methods taking visual input only.}
\label{tab:diffv2s_secs}
\end{table}

\paragraph{Audio-Visual Synchronization}
Lip-Sync Expert (LSE) \citep{lipsyncexpert_lsemetric} provides a subjective evaluation of audio-visual synchronization, making it useful for assessing the alignment between synthesized speech and the original video. Higher LSE-C values and lower LSE-D values indicate better synchronization. LSE metrics are calculated following the implementation of \citet{lipsyncexpert_lsemetric, syncnetChung16a}.
Table \ref{tab:lse} presents the LSE-C and LSE-D evaluation results, which reflect the model's ability in audio-visual synchronization in LRS3 dataset. We compared DiVISe against existing methods on LSE-C and LSE-D scores and have found that DiVISe has achieved superior LSE scores even better than methods requiring speaker embedding as input \cite{mira2022svts, Kim2023MultiTaskL2S}.
DiVISe shows a prominent performance in audio-visual synchronization with either Griffin-Lim or HiFi-GAN vocoder as means of Mel-to-Audio conversion. This elevates lip-sync confidence score suggests that the audio track produced by our method synchronizes well with the speaker lip movements in video, signifying a strong audio-visual correlation.

\begin{table}[t]
\centering
\small
\begin{tabular}{lcc}
\toprule
\textbf{Method}&\textbf{LSE-C}$\uparrow$&\textbf{LSE-D}$\downarrow$\\
\hline
\multicolumn{3}{c}{\textit{Methods requiring extra information}} \\
SVTS & 6.04\dag & 8.28\dag \\
Multi-Task & 5.19\dag & 8.89\dag \\
\hline
\multicolumn{3}{c}{\textit{Methods requiring visual input only}}
\\
DiffV2S & 7.28\dag & 7.27\dag \\
ReVISE & 7.11 & 7.20 \\
DiVISe (Griffin-Lim) & 7.35 & 6.90 \\
DiVISe (Vocoder) & \textbf{7.85} & \textbf{6.52} \\
\bottomrule
\end{tabular}%
\caption{LSE-C and LSE-D results for audio-visual synchronization evaluation. Results reproduced by \citet{choi2023diffv2s} are marked with \dag.}
\label{tab:lse}
\end{table}

\paragraph{Human Subjective Speaker Matching Score}\label{sec:subjective_speaker_matching}

\begin{table}[t]
\centering
\small
\begin{tabular}{lc}
\toprule
\textbf{Audio Source}&\textbf{Matching Score}\\
\hline
\multicolumn{2}{c}{\textit{Vocoders}}\\
Unit-HiFiGAN & 1.81±0.18 \\
HiFi-GAN & 4.64±0.08 \\
\hline
\multicolumn{2}{c}{\textit{V2S Methods}}\\
ReVISE & 1.80±0.18 \\
DiVISe & 3.90±0.14 \\
\hline
\multicolumn{2}{c}{\textit{Ground Truth}}\\
Source & \textbf{4.80±0.07} \\
\bottomrule
\end{tabular}%
\caption{Subjective measurements of Speech Matching Scores, where the relation of synthesized audio to the physical appearance of speaker in the original video is assessed by test participants.}
\label{tab:speaker_matching_score}
\end{table}

In this section, we conducted a subjective human evaluation on speaker matching, where participants were required to rate the synthesized speech based on its similarity to the speaker's image. We encourage readers to refer to Section \ref{sec:rs_subjective_speaker_matching} for more details about this test. The evaluation results are reported in Table \ref{tab:speaker_matching_score}, with a confidence interval of 95\%.
The speaker matching scores for mel-based methods significantly outperform those of unit-based methods, indicating greater human satisfaction with the audio-visual connection in speaker characteristics. DiVISe excels in aligning the generated audio with the physical characteristics of the visual speaker's appearance. This framework demonstrates considerable potential in fusing acoustic features from the input video, which are essential for accurately restoring speaker characteristics in the synthesis.

\subsection{Results on Intelligibility}\label{sec:results}

In this section, we presented the synthesis results for intelligibility in the full resource setting. It is important to highlight that some existing methods incorporate additional information, such as speaker embeddings or text transcriptions, which gives them an advantage over DiVISe.

Despite this disparity, DiVISe demonstrates remarkable performance improvements, as detailed in Table \ref{tab:synthesis_results}. Specifically, it demonstrates superior speech content recovery across both datasets in WER, outperforming all other approaches on LRS2 and showing competitive performance on LRS3. Additionally, DiVISe delivers the highest STOI and ESTOI scores in this setting, indicating superior audio quality and intelligibility. These results underscore DiVISe’s robustness, as it performs exceptionally well without requiring additional data. Moreover, DiVISe consistently demonstrates generalizability, particularly on LRS2, which is not part of the AV-HuBERT backbone's pre-training corpus.

\subsection{Ablation Studies}

In this subsection, we conduct ablation studies on model composition and resource settings.

\subsubsection{Analysis on Size of Data and Model}

\begin{figure}[t]
    \centering
    \includegraphics[width=\linewidth]{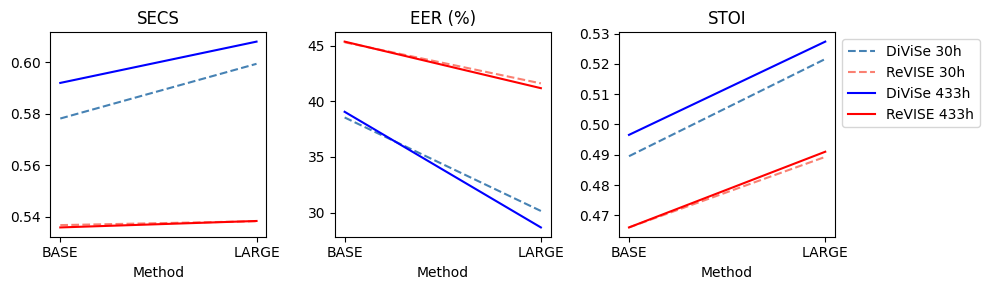}
    \caption{Comparative evaluation results with different model sizes on LRS3.}
    \label{fig:synthesis_results_size_varied}
\end{figure}

In this section, we analyzed model performance in speaker characteristics preservation and intelligibility under varying model size and training resource. The corresponding visualizations of these results are presented in Figure \ref{fig:synthesis_results_size_varied}.

\paragraph{Size of data} DiVISe demonstrates superior performance across all metrics in the full resource setting (433 hours) compared to the low resource setting (30 hours), consistently outperforming ReVISE in speaker matching. As shown in Table \ref{tab:synthesis_results}, despite having a smaller training corpus, DiVISe exceeds most reported methods in intelligibility, even when those methods are trained on larger datasets. Moreover, compared to ReVISE, DiVISe's performance scales up more profoundly with more data, which suggests that DiVISe has significant potential for scaling effectively to larger datasets.

\paragraph{Size of model} DiVISe benefits significantly from increased model parameters. In Figure \ref{fig:synthesis_results_size_varied}, we compare the performance of AV-HuBERT BASE and LARGE (the default setup in our work) as visual backbones, with LARGE being three times larger than BASE. Compared to ReVISE, DiVISe shows more substantial performance improvements across all metrics as the model size increases. This finding suggests that DiVISe achieves greater performance gains as it scales to larger models, making it well-suited for broader V2S applications that require support for a larger number of model parameters.

\subsubsection{Effect of Audio-Visual Pre-training}

\begin{table}[t]
\centering
\small
\setlength{\tabcolsep}{1mm}{
\begin{tabular}{lccccc}
\toprule
\multirow{2}{*}{\textbf{Pre-train}} & \multirow{2}{*}{\textbf{Method}} & \multicolumn{2}{c}{Match} & \multicolumn{2}{c}{Intell.} \\
&&\textbf{SECS}$\uparrow$&\textbf{EER}$\downarrow$&\textbf{ESTOI}$\uparrow$&\textbf{WER}$\downarrow$\\
\hline
\checkmark & DiVISe & \textbf{0.6242} & \textbf{28.66} & \textbf{0.331} & \textbf{35.68} \\
$\times$ & DiVISe & 0.4842 & 37.62 & 0.206 & 93.98 \\

\checkmark & ReVISE & 0.5384 & 41.17 & 0.290 & 36.03 \\
$\times$ & ReVISE & 0.5304 & 44.25 & 0.220 & 77.24 \\
\bottomrule
\end{tabular}%
}
\caption{Ablation on audio-visual pre-training.}
\label{tab:ablate_pretrain}
\end{table}

Pre-training is essential for both DiVISe and ReVISE, as shown by the performance gains in Table \ref{tab:ablate_pretrain} with pre-trained AV-HuBERT.
Pre-training results in significant improvements across all metrics for DiVISe, while ReVISE primarily benefits in intelligibility but fails to meaningfully improve speaker preservation. This disparity underscores ReVISE's limitations in retaining speaker characteristics, which remain unrecovered despite the knowledge acquired from pre-training.

\subsubsection{Effect of Conformer Module}\label{sec:conformer_module}

\begin{table}[t]
\centering
\small
\begin{tabular}{ccccccc}
\toprule
\multirow{2}{*}{\textbf{Method}} & \multicolumn{2}{c}{Match} & \multicolumn{2}{c}{Intell.} \\
&\textbf{SECS}$\uparrow$&\textbf{EER}$\downarrow$&\textbf{ESTOI}$\uparrow$&\textbf{WER}$\downarrow$\\
\hline
\makecell{DiVISe} & \textbf{0.6242} & \textbf{28.66} & \textbf{0.331} & \underline{35.68} \\
\makecell{- Conformer} & \underline{0.6130} & \underline{29.72} & \underline{0.324} & 38.41 \\ 
\makecell{ReVISE} & 0.5384 & 41.17 & 0.290 & 36.03 \\
\makecell{+ Conformer} & 0.5393 & 41.69 & 0.291 & \textbf{35.67} \\ 
\bottomrule
\end{tabular}%
\caption{Ablation on Conformer module.}
\label{tab:ablate_conformer}
\end{table}

As shown in Table \ref{tab:ablate_conformer}, integrating the conformer module into DiVISe improves acoustic intelligibility, demonstrating its ability to analyze temporal visual dynamics. In contrast, ReVISE \cite{revisewnhu} shows more limited improvement with the conformer. For both methods, the integration mainly boosts intelligibility, with little effect on preserving speaker characteristics.

\subsubsection{Effect of Vocoder Fine-tuning}

\begin{table}[t]
\centering
\small
\begin{tabular}{ccccccc}
\toprule
\multirow{2}{*}{\textbf{Fine-tune}} & \multicolumn{2}{c}{Match} & \multicolumn{2}{c}{Intell.} \\
&\textbf{SECS}$\uparrow$&\textbf{EER}$\downarrow$&\textbf{ESTOI}$\uparrow$&\textbf{WER}$\downarrow$\\
\hline
\makecell{\checkmark} & \textbf{0.6242} & \textbf{28.66} & \textbf{0.331} & \textbf{35.68} \\
\makecell{$\times$} & 0.5505 & 32.32 & 0.330 & 36.26 \\
\bottomrule
\end{tabular}%
\caption{Ablation on vocoder fine-tuning for DiVISe.}
\label{tab:ablate_vc_finetune}
\end{table}

Table \ref{tab:ablate_vc_finetune} highlights the effect of fine-tuning the vocoder on V2S frontend outputs. Fine-tuning leads to significant improvements, particularly in speaker matching scores, underscoring its importance for DiVISe. In contrast, intelligibility metrics show minimal impact from fine-tuning, likely due to their robustness to slight frequency variations.

\section{Latency and Computational Efficiency}

We conduct analysis of latency and computational efficiency in Table \ref{tab:latency_comparison}. For latency, we conducted inference on the LRS3 test set and reported the average time consumption per sample and per video frame. For computational efficiency, we calculated FLOPs for a 4-second input video, with all evaluations performed on a single RTX4090 GPU.

\begin{table}[ht]
\centering
\small
\begin{tabular}{l|cccc}
\toprule
\multirow{2}{*}{\textbf{Component}} & \multirow{2}{*}{\textbf{Method}} & \multicolumn{1}{c}{Latency} & \multicolumn{2}{c}{Throughput (/s)} \\
&  & \textbf{FLOPs} & \textbf{samples} & \textbf{frames} \\ \hline
Frontend & DiVISe* & 112.70 & \textbf{37.07} & \textbf{2127} \\
Full Model & DiVISe* & 327.54 & 15.76 & 904 \\ \hline
Frontend & DiVISe & 121.58 & 27.81 & 1595 \\
Full Model & DiVISe & 336.42 & 13.73 & 788 \\ \hline
Frontend & ReVISE & 115.96 & 35.69 & 2047 \\
Full Model & ReVISE & 251.90 & 15.83 & 908 \\ \hline
Frontend & SVTS & \textbf{85.06} & 24.54 & 1408 \\
Full Model & SVTS & 299.90 & 12.45 & 714 \\
\bottomrule
\end{tabular}
\caption{Comparison of FLOPs and latency. Full Model means a comination of V2S frontend and vocoder. DiVISe* refers to DiVISe w/o Conformer.}
\label{tab:latency_comparison}
\end{table}

We compare DiVISe with ReVISE \citep{revisewnhu} and SVTS-L \citep{mira2022svts}. Our results show that the model achieves moderate computational efficiency overall. Notably, when the conformer module is disabled as studied in Section \ref{sec:conformer_module}, our model achieves higher throughput than \citet{revisewnhu}. We summarize the key points to note as follows:

\paragraph{Vocoder Processing Rates} Unit-HiFiGAN operates on HuBERT units at 50 Hz, while our HiFi-GAN implementation uses a hop size of 160, corresponding to 100 Hz (16k/160, where 16k is the audio sampling rate). This results in the vocoder input sequence length for HiFi-GAN and DiVISe being twice that of ReVISE, leading to higher FLOPs. To illustrate, we compared the FLOPs for vocoder inputs extracted from 1-second audio in Table \ref{tab:flops_vocoder}.

\begin{table}[ht]
    \centering
    \small
    \begin{tabular}{lcc}
    \toprule
        \textbf{Vocoder} & \textbf{FLOPs} & \textbf{Input Rate} \\\hline
        HiFi-GAN&53.72&100/s \\
        Unit-HiFiGAN&33.98&50/s \\ 
    \bottomrule
    \end{tabular}
    \caption{Comparison of FLOPs for vocoder inputs.}
    \label{tab:flops_vocoder}
\end{table}

\paragraph{Frontend Throughput} Disabling the conformer module in our V2S frontend improves DiVISe's throughput beyond that of ReVISE (both in frontend and overall performance). This improvement is primarily due to the linear reshaping upsampling approach described in Section \ref{sec:model_architecture}, which eliminates the need for transpose convolution used in the ReVISE frontend. The difference in vocoder processing rates does not significantly affect these results.

\paragraph{Comparison with SVTS-L} Although SVTS-L is slightly smaller in model size (87.6M reported in \citet{mira2022svts}), it requires a speaker encoder that processes additional audio input, introducing extra latency. In contrast, both DiVISe and ReVISE rely solely on silent video inputs, resulting in superior throughput compared to SVTS-L.

\section{Conclusion}

In this study, we introduced DiVISe, a V2S model directly predicting Mel-spectrograms from silent videos without relying on any additional inputs both training and inference. By leveraging mel-based vocoder and audio-visual pre-training,
DiVISe achieves superior results in speaker characteristics preservation, as confirmed through both subjective and objective evaluations. Besides, DiVISe enhances speech intelligibility, accurately recovering the audible content of speech.
Additionally, our findings indicate that DiVISe scales more efficiently with larger datasets and model sizes.

Overall, DiVISe delivers high-quality, intelligible speech with minimal input requirements, highlighting its potential for broader applications in speech synthesis and related fields. In the future, DiVISe can be applied to larger datasets and enhanced with increased model parameters.

\section{Limitations}

Following \citet{Shi2022AVHuBERT}, our current work requires preprocessing the input frames as detailed in Section \ref{sec:preprocessing} of appendix. This limits the application of DiVISe when applied in real-time V2S scenario, especially when locating the area of interest centered around the mouth for each input video frame. This could lead to extra lag in real-time application, and we expect potential streamlining in current preprocessing pipeline to make it more suited towards real-time V2S synthesis in the future.

DiVISe is only tested on LRS2 and LRS3 datasets containing English-only corpus. It is not tested on multi-lingual audio-visual corpus in this work due to the limitation of computational resources, but we still expect to carry out future research on its capability of adapting to other languages (e.g. Chinese, French).

\bibliography{custom}

\appendix

\section{Training Details}\label{sec:training_details}

\subsection{Preprocessing Details}\label{sec:preprocessing}

DiVISe has adopted the data processing pipeline from AV-HuBERT \cite{Shi2022AVHuBERT} for data consistency as the upstream model is a pre-trained AV-HuBERT model. The area of interest, specifically a 96×96 pixel region centered around the mouth, is then extracted from each frame\footnote{We applied the dlib tool \cite{dlibToolKing2009} to each video clip in order to detect the 68 facial keypoints. Subsequently, we employ an affine transformation to align each video frame with a standard reference face frame.}. In training, we further process the images by extracting a smaller 88×88 region from the larger mouth-centered area. The selected frames are subject to a 50\% chance of being horizontally flipped to enhance the robustness of our model. In training we limit the length of each sampled video clip to 4.0 seconds starting from the beginning of the video. At inference, the entire video is always loaded into the model, except for EER evaluation. We consistently perform a center crop of the 88×88 region without applying any horizontal flipping in evaluation.

\subsection{Parameters Setting}\label{sec:params_setting}
For LRS2 and LRS3 dataset, $h_0$ is set to 25 to match the sampling frequency and the final waveform is at a frequency of 16kHz. We follow the established practice of \citet{Niekerk2021ACOSoftSpeechUnitsCompare} by extracting the 128-dimensional Mel-Spectrogram ($F=128$) from the raw audio waveform at 10-ms intervals for both the LRS3 and LJSpeech datasets.

We choose a hop size that aligns $\mathcal{S}$ with $\tilde{\mathcal{S}}$ at the same sampling frequency ($kh_0$). Hop size is set to be 160 to produce Mel-spectrogram at 100Hz. In this way, the upsampled visual features can match with the sampling frequency of the Mel-spectrogram ($k\times25=100$Hz), where upsamling ratio $k=4$.

\subsection{Training And Evaluation Sets}

We follow the setup of AV-HuBERT \cite{Shi2022AVHuBERT} to determine different sets of datasets in hours. The LRS2 is partitioned into three sections (pre-train for 195 hours, trainval for approximately 30 hours, and test for 0.5 hours), with a predefined validation set. The LRS3 dataset is segmented into three splits (pre-train for around 400 hours, trainval for 30 hours, and test for 1 hour), with the validation set being extracted from the same subset of the LRS3 trainval split as identified by \citet{Shi2022AVHuBERT}.

The \lrsbasesetting setting employs only the trainval split (excluding the validation set) from the LRS2 and LRS3 dataset for model training, while the \lrsfullsetting and \lrstwofullsetting setting combine pre-train and trainval splits for LRS3 and LRS2, respectively. Regardless of resource setting, the validation and test sets remain consistent for each dataset in evaluation, with the sole exception of EER evaluation, which only uses test pairs of trials for each item in the VoxCeleb2 test set specified in Section \ref{sec:metrics}.

\subsection{Experimental Configurations}\label{sec:settings}


\subsubsection{V2S frontend}

In our experimental setup, DiVISe leverages the AV-HuBERT LARGE model \cite{Shi2022AVHuBERT}, which is pre-trained for 5 epochs on the LRS3 dataset and the English-only subset of VoxCeleb2 \cite{VoxCeleb2Chung_2018}, comprising a total of 325 million parameters\footnote{The pre-trained weights used in this work are publicly accessible at \url{https://facebookresearch.github.io/av_hubert/}}.
This choice aligns with the setting of \citet{revisewnhu} and serves as the default configuration for the experiment. For comparison, the AV-HuBERT BASE model \cite{Shi2022AVHuBERT} with 103 million parameters, is pre-trained solely on LRS3. This yields a relatively modest enhancement for our method. The conformer module is modified from the implementation of \citet{Ma_2022_VSRMultiLanguageIntheWild}, with the visual frontend removed.

For both ReVISE and DiVISe, V2S frontend training was accomplished on 8 GPUs, with each GPU requiring fewer than 45K updates for about 26 hours. Models are selected based on the lowest L1 loss of Mel-spectrogram and highest accuracy of unit prediction for DiVISe and ReVISE respectively.
Specific configurations on LRS2 and LRS3 dataset can be seen in Table \ref{tab:appendix_setting_general} and Table \ref{tab:appendix_setting_resource}. Table \ref{tab:appendix_setting_general} exhibits full resource setting and low resource setting, with their differences specified in Table \ref{tab:appendix_setting_resource}. The total number of updates may be slightly less than those detailed in Table \ref{tab:appendix_setting_resource}, due to the model selection criteria that we applied. We follow the experimental settings of \citet{revisewnhu} for the full resource setting, but reduce the number of total updates to 1/4 and the number of GPUs to 1/2 to avoid overfitting the volume of data in the low resource setting. The tri-stage LR schedule refers to the learning rate schedule that increases linearly from 0 in the first stage, constantly staying at the maximum learning rate value in the second stage, and finally linearly decreasing to 5\% of the peak learning rate value in the last stage. The LR schedule evolves on the per-update level. We freeze the pre-trained AV-HuBERT module for a few updates at the beginning of training.
\begin{table}[h]
\centering
\begin{tabular}{l|c}
\toprule
tri-stage LR schedule & (10\%, 20\%, 70\%) \\
peak learning rate & 6e-5 \\
batch size / GPU & 10 \\
max sample length & 4s \\
Adam \((\beta_1, \beta_2)\) & (0.9, 0.98) \\
\bottomrule
\end{tabular}
\caption{Experimental settings for LRS2 and LRS3.}
\label{tab:appendix_setting_general}
\end{table}

\begin{table}[h]
\centering
\begin{tabular}{l|cc}
\toprule
\textbf{Settings of Resource}&\textbf{Full}&\textbf{Low}\\
\hline
num. of updates & 45000 & 11250 \\
num. of frozen steps & 5000 & 1250 \\
num. of GPU & 8 & 4 \\
\bottomrule
\end{tabular}
\caption{Experimental settings for different resources.}
\label{tab:appendix_setting_resource}
\end{table}

\subsubsection{Vocoder}

For DiVISe, we trained HiFi-GAN with 128 Mel-frequency bins as the neural vocoder to ensure consistency with the ReVISE framework \cite{revisewnhu}. HiFi-GAN is first trained on the resampled LJSpeech dataset for 400k updates with a single GPU.
We then fine-tune HiFi-GAN on Mel-spectrograms generated by upstream V2S frontend trained on full-resource setting of LRS3 and LRS2 datasets with 8 GPUs, respectively. We apply the best NISQA \citep{Mittag_2021_NISQAMOS} prediction score as model selection criterion for acoustic perceptual quality\footnote{We use nisqalib available at \url{https://github.com/kale4eat/nisqalib} to obtain NISQA scores.}. Vocoders selected for audio synthesis is given in Table \ref{tab:appendix_vocoder_finetuning}.
The training code is developed based on the implementation of \citet{Niekerk2021ACOSoftSpeechUnitsCompare}.

\begin{table}[h]
\centering
\begin{tabular}{l|l}
\toprule
\textbf{LRS3}&\textbf{LRS2}\\
\hline
30k & 50k\\
\bottomrule
\end{tabular}
\caption{Number of updates for vocoder finetuning selected given best validation NISQA scores.}
\label{tab:appendix_vocoder_finetuning}
\end{table}

For ReVISE implementation, we align with the settings of \citet{revisewnhu}. We trained a 2000-class K-Means model on LRS3 from feature extracted in the last layer of the HuBERT BASE model pre-trained on LibriSpeech 960h dataset \citep{LibrispeechLS960} in the third epoch\footnote{The pre-trained HuBERT model used for replicating ReVISE in this study is obtained from \url{https://dl.fbaipublicfiles.com/hubert/hubert_base_ls960.pt}}. We then train Unit-HiFiGAN on units labeled by this K-Means model for 400k updates on the resampled LJSpeech dataset with 8 GPUs. Since the unit vocabulary is shared with its upstream V2S frontend in prediction, there is no need for further fine-tuning.

\section{Model Sizes}\label{sec:model_layout}
\begin{table}[t]
\centering
\small
\begin{tabular}{lccc}
\toprule
\textbf{Modules} & \textbf{DiVISe} & \textbf{DiVISe-BASE} & \textbf{ReVISE}\\
\hline
Visual Backbone & 325M & 103M & 325M \\
Upsampling & - & - & 8M \\
Conformer & 11M & 11M & - \\
\hline
Sum w/o Vocoder & 336M & 114M & 333M \\
\hline
Vocoder & 14M & 14M & 15M \\
\hline
Sum w/ Vocoder & 350M & 128M & 348M \\
\bottomrule
\end{tabular}
\caption{Parameters comparison of DiVISe in model size. DiVISe-BASE replaces the visual backbone with pre-trained AV-HuBERT BASE model, trimming model size to 1/3.}
\label{tab:model_size}
\end{table}
From Table \ref{tab:model_size}, one can see that the inclusion of the conformer module in our proposed model adds merely 10.29 million parameters, which is relatively minimal when compared to the overall size of the model. This demonstrates that the conformer module does not significantly increase the model's size. 
DiVISe worked around the upsampling module by directly applying linear reshaping following \citet{Shi2022AVHuBERT}. Substituting the visual backbone with AV-HuBERT BASE results in an even more compact version of the proposed model (DiVISe-BASE), suitable for scenarios requiring a lighter computational footprint. In comparision, DiVISe maintains a relatively moderate size.


\section{Subjective Mean Opinion Score Evaluation}\label{sec:subjective_mos}

\begin{table}[ht]
\centering
\small
\begin{tabular}{lc}
\toprule
\textbf{Audio Source}&\textbf{MOS Score}\\
\hline
Unit-HiFiGAN & $4.37\pm0.11$\\
HiFi-GAN & $4.24\pm0.12$ \\
Ground Truth & \textbf{$4.47\pm0.11$} \\
\bottomrule
\end{tabular}%
\caption{Subjective MOS score measurements for vocoders on LRS3 dataset.}
\label{tab:mos_score}
\end{table}

We conducted a subjective evaluation for Unit-HiFiGAN and HiFi-GAN to further explore subjective human feedback on their synthesized audio clips. The confidence interval for Mean opinion score (MOS) scores is 95\%. Other settings can be found in Section \ref{sec:rs_subjective_speaker_matching}. In Table \ref{tab:mos_score}, there is a slight drop in the MOS score for HiFi-GAN compared to Unit-HiFiGAN. We assume that this performance gap is attributed to the relatively more consistent and neutral style of audio clips in all samples generated by unit-based methods, without considering the identity of the speaker. Our experiment reveals that this trait of unit vocoder gains a better impression from human listeners. However, it is important to note that no speaker-related information is given to test participants to assess how related synthesized audio clips are to the visual speaker, which is an in-born drawback for this audio-only MOS evaluation.

\section{More Details on Metrics}\label{sec:metrics}

In this section, we provide detailed explanation and implementation for metrics mentioned in this work for reader's better understanding.

\paragraph{Speech Similarity and Speaker Verification} For speech similarity, SECS is computed by simply averaging the cosine similarity between speaker embeddings of ground-truth and synthesized audio \citep{choi2023diffv2s}. On the other hand, EER is a metric used to assess the performance of speaker verification. For EER evaluation, we first set up a list of positive and negative pairs for every sample in the VoxCeleb2 test set that sums up to a total of 72,474 pairs of trials, similar to \citet{Shi2022LearningAudioVisualSpeakerEmbedding_SCSV}. To compute EER, we first synthesize audio paris with V2S methods given any video pair of trials, then we use the cosine similarity of speaker embeddings computed from the generated audio clips for every pair of trials as the matching score. The matching score serves as the prediction score, where ground-truth labels are assigned as 1 for positive pairs and 0 for negative pairs. These scores are used to compute false acceptance rates and false rejection rates. To simplify EER calculation, we use only one pair of 4-second audio clips for each pair of trials. For both SECS and EER, we use an existing pre-trained audio speaker encoder \cite{jia2019transfer} to gain speaker embedding in a setup identical to \citet{choi2023diffv2s}.

\paragraph{Intelligibility} The intelligibility of synthesized audio clips is quantitatively assessed by objective metrics, specifically Short-Time Objective Intelligibility (STOI), Extended STOI (ESTOI) and Perceptual Evaluation of Speech Quality (PESQ)\footnote{The PESQ metric was calculated using version 0.0.4 of the pesq Python package, whereas STOI and ESTOI metrics were determined with version 0.3.3 of the pystoi Python package.}. These metrics provide a comprehensive understanding of the speech intelligibility and quality that our model is capable of producing. 


Additionally, the synthesized waveforms are further transcribed into text using the publicly available wav2vec 2.0 model pre-trained by \citet{Xu2020SelfTrainingAP_reviseASR}\footnote{The wav2vec 2.0 model with its weights used for ASR evaluation is sourced from \url{https://huggingface.co/facebook/wav2vec2-large-960h-lv60-self}} identical to the choice of ReVISE \cite{revisewnhu}. This ASR system allows us to evaluate the intelligibility of the generated speech in terms of its WER (the lower the better) with the text transcribed from ground-truth audio. The transcription results also serve as a proxy for how well the model's output would be understood in real-world applications, where accurate conversion of speech to text is often crucial. This ASR reports a WER of 5. 68\% in the LRS3 dataset in our setting.

\section{Performance on various text lengths}
\label{sec:qualitative Examples}
\begin{figure}[ht]
    \centering
    \includegraphics[width=\linewidth]{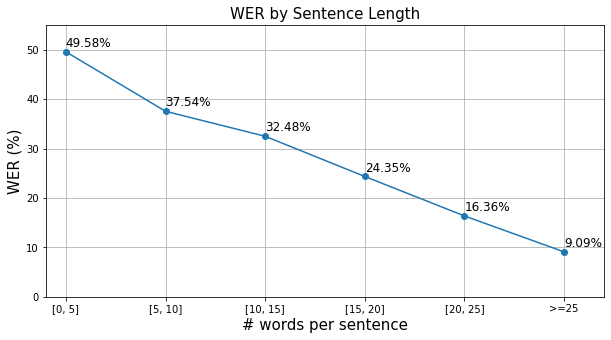}
    \caption{WER reported on different length spans of Ground-Truth texts. Video clips corresponding to longer lengths generally have better WER performance for DiVISe.}
    \label{fig:wer_by_sentence_length}
\end{figure}
The graph in Figure \ref{fig:wer_by_sentence_length} presents the WER by sentence length of the text transcribed from ground-truth audio for DiVISe. DiVISe exhibits a decrease in WER as the length of the sentence increases. A similar observation is also found by \citet{Shi2022AVHuBERT}, where WER decreases as the length of ground-truth text increases in lip-reading (VSR) task. DiVISe gains better performance at handling longer sequences of words evaluated by ASR in the V2S setting, possibly due to the availability of more contextual information within the video that aids in speech synthesis. This could mean that, intuitively, V2S and VSR are two highly correlated tasks that share similar traits.

\section{Research Ethics and Volunteer Testing}

\subsection{Use of Artifacts}

For all datasets used in this work (LRS3-TED \citep{Afouras2018LRS3TEDAL}, LRS2-BBC \citep{LRS2Afouras_2022}, VoxCeleb2 \citep{VoxCeleb2Chung_2018}, and LJSpeech \citep{ljspeech17}), we strictly adhere to the usage guidelines outlined by their respective authors. The datasets we use are publicly available online and do not contain any personally identifying information or offensive content. No additional steps were required for anonymization or content screening, as the data has already been curated and verified for research purposes by the original providers. No other data has been collected, repackaged, or released as part of this work.

For the code artifacts used in this work, all relevant code has been appropriately credited through proper citations. Metrics, results, and other contributions from external authors have been thoroughly acknowledged and referenced. The use of existing artifacts in this work aligns with their intended purpose as specified by their original creators. For the artifacts we generate, their intended use is explicitly for research purposes.

\subsection{Human Subjects Setup in Subjective Evaluation}\label{sec:rs_subjective_speaker_matching}
In this study, we recruited 15 student participants to conduct subjective evaluations of the synthesized speech data. Participation was voluntary, with no financial compensation or course credit provided. Each participant received detailed instructions, including a step-by-step guide and screenshots of the evaluation interface, ensuring clarity in the evaluation process. Since the study posed no potential risks to the participants, no specific disclaimers were necessary. Informed consent was obtained prior to participation, with students agreeing to the use of their anonymized evaluation data solely for research purposes. Additionally, participants were explicitly informed that no personal identifying information (PII) would be collected or stored during the study.

In Section \ref{sec:subjective_speaker_matching}, we performed a subjective speaker matching evaluation, in which each test participant is provided with a head-cropped picture of the speaker in the original video and the corresponding ground-truth speech as reference. Based on the given reference, the participants are asked to give their opinion scores on a list of candidate speeches synthesized from difference sources. This subjective evaluation was carried out on 20 randomly selected samples from the LRS3 test set. The full text of instruction given to the test participant is:
\begin{quote}
    Thank you for your time to take this test! In the following tests, you will be given a speaker image paired with different speech clips. You are only required to score (from 1 to 5) how much you think the given speech matches the speaker image. Speech content and quality do not matter in this test.
    
    We recommend you to put on your earphone for better intelligibility of audio.
\end{quote}

In Section \ref{sec:subjective_mos}, we also performed subjective MOS test on vocoders on the same range of samples. The test settings align with the setup of Section \ref{sec:subjective_speaker_matching}. As vocoders should be evaluated in an audio-only setup, no reference speaker image is provided. The full text of instruction is:

\begin{quote}
    Thank you for your time to take this test! In the following tests, you will be given different speech clips. You are only required to score (from 1 to 5) on the quality of the speech. Slight differences in speech content compared to the reference are normal and can be disregarded.
    
    We recommend you to put on your earphone for better intelligibility of audio.
\end{quote}

\section{Transcription Examples}

\begin{table*}[ht]
\centering
\begin{tabularx}{\linewidth}{>{\raggedright\arraybackslash}p{0.2\linewidth}|>{\raggedright\arraybackslash}X}

    \toprule
    \textbf{GT} & We're on the grid \\
    \textbf{ASR (GT)} & \textcolor{red}{Were} on the grid \\
    \textbf{ASR (DiVISe)} & \textcolor{red}{Were} on the \textcolor{red}{green} \\
    \hline

    \textbf{GT} & I was not going to cry myself to sleep \\
    \textbf{ASR (GT)} & I was not going to cry myself to sleep \\
    \textbf{ASR (DiVISe)} & I was not \textcolor{red}{in} to \textcolor{red}{crow} myself \textcolor{red}{asleep} \\
    \hline

    \textbf{GT} & One woman no longer believes love will ever find her \\
    \textbf{ASR (GT)} & One woman no longer believes love will ever find her \\
    \textbf{ASR (DiVISe)} & \textcolor{red}{Where} woman no longer believes \textcolor{red}{lovel} ever \textcolor{red}{factor} \\
    \hline

    \textbf{GT} & I think we need something like a Manhattan Project on the topic of artificial intelligence \\
    \textbf{ASR (GT)} & I think we need something like a Manhattan Project on the topic of artificial intelligence \\
    \textbf{ASR (DiVISe)} & \textcolor{red}{Ethic} \textcolor{red}{would} \textcolor{red}{lead} something like a Manhattan Project on the topic of artificial \textcolor{red}{case} \\
    \hline

    \textbf{GT} & And he was talking about the importance of coaching boys into men and changing the culture of the locker room and giving \\
    \textbf{ASR (GT)} & And he was talking about the importance of coaching boys into men and changing the culture of the locker room and giving \\
    \textbf{ASR (DiVISe)} & And he was talking about the importance of coaching \textcolor{red}{poisons} \textcolor{red}{a} \textcolor{red}{mine} and changing the culture of the \textcolor{red}{lakarone} and \textcolor{red}{even} \\
    \hline

    \textbf{GT} & So people hear about this study and they're like great if I want to get better at my job I just need to upgrade my browser \\
    \textbf{ASR (GT)} & So people hear about this study and \textcolor{red}{their} \textcolor{red}{leg} \textcolor{red}{greate} if I want to get better at my job I just need to upgrade my browser \\
    \textbf{ASR (DiVISe)} & So people hear about \textcolor{red}{the} study and they're like \textcolor{red}{radophoni} to get better my job I just \textcolor{red}{said} to \textcolor{red}{operate} my \textcolor{red}{brother} \\
    \bottomrule
\end{tabularx}
\caption{DiVISe has the ability to create audio that yields audio-to-text conversion outcomes similar to ground truth when transcribed by ASR. Differences between transcribed text and ground-truth text are marked in red.}
\label{tab:appendix_qualitative}
\end{table*}

This section evaluates the intelligibility of the audio generated by our method through a comparative analysis with ground-truth audio transcribed by the ASR system. Table \ref{tab:appendix_qualitative} features examples of transcriptions from ground-truth audio and our generated audio, as interpreted by the ASR system.

The examples showcase the effectiveness of DiVISe in accurately replicating ground-truth transcriptions through its audio output. For example, "\textit{were on the green}" closely resembles the ASR transcription "\textit{were on the grid}" from ground-truth audio, demonstrating the high intelligibility of the generated audio. However, there are also notable discrepancies due to unclear articulation in the generated audio, such as "\textit{I think we need}" being interpreted as "\textit{Ethic would lead}". These examples highlight the strengths of DiVISe in generating intelligible audio that can be effectively transcribed by an ASR system.

\end{document}